\documentclass{pasj00}

\begin{document}
\SetRunningHead{Author(s) in page-head}{Running Head}
\Received{2009/04/12}
\Accepted{2009/06/30}

\title{Implications of Bulk Velocity Structures in AGN Jets}

\author{Jianping \textsc{Yang} %
  \thanks{National Astronomical Observatories, Yunnan Observatory,
Chinese Academy of Sciences,  Kunming 650011, China}}
\affil{National Astronomical Observatories, Yunnan Observatory,
Chinese Academy of Sciences,  Kunming 650011, China; Graduate School
of the Chinese Academy of Sciences, China; Yunnan Agricultural
University, Kunming 650201, China} \email{yangjp@mail.ynao.ac.cn}

\author{Jiancheng \textsc{Wang}}
\affil{National Astronomical Observatories, Yunnan Observatory,
Chinese Academy of Sciences,  Kunming 650011,
China}
\author{Benzhong {\sc Dai}}
\affil{Department of Physics, Yunnan University, Kunming 650091,
China}
\author{Xiaoyan {\sc Gao}} \affil{National Astronomical Observatories, Yunnan Observatory,
Chinese Academy of Sciences,  Kunming 650011, China}

%


\maketitle

\begin{abstract}
The synchrotron self-Compton (SSC) models and External Compton (EC)
models of AGN jets with continually longitudinal and transverse bulk
velocity structures are constructed. The observed spectra show
complex and interesting patterns in different velocity structures
and viewing angles. These models are used to calculate the
synchrotron and inverse Compton spectra of two typical BL Lac
objects (BLO) (Mrk 421 and 0716+714) and one Flat Spectrum Radio
Quasars (FSRQs) (3c 279), and to discuss the implications of jet
bulk velocity structures in unification of the BLO and FR I radio
galaxies (FRI). By calculating the synchrotron spectra and SSC
spectra of BL Lac object jets with continually bulk velocity
structures, we find that the spectra are much different from ones in
jets with uniform velocity structure under the increase of viewing
angles. The unification of BLO and FRI is less constrained by
viewing angles and would be imprinted by velocity structures
intrinsic to the jet themselves. By considering the jets with bulk
velocity structures constrained by apparent speed, we discuss the
velocity structures imprinted on the observed spectra for different
viewing angles. We find that the spectra are greatly impacted by
longitudinal velocity structures, becasue the volume elements are
compressed or expanded. Finally, we present the EC spectra of FSRQs
and FR II radio galaxies (FRII) and find that they are weakly
affected by velocity structures compared to synchrotron and SSC
spectra.

\end{abstract}

\section{Introduction}

It is well known that the AGN jets radiate the strong nonthermal
emission from the radio to the gamma ray range. Their spectral
energy distribution (SED) consists of two bumps, attributed to the
synchrotron and the inverse Compton (IC) emission of
ultrarelativistic particles. The simple one-zone homogeneous jet
model usually provides a good framework to explain the emissive
spectra of AGN jets. However, the realistic case is that the jet has
the bulk velocity structures in longitudinal and transverse
directions to its axis \citep{Blandford95}. To explain the spectral
properties of some AGNs several authors have proposed many
inhomogeneous jet models \citep{Marscher80, Ghisellini85,
Georganopoulos98, Wang04,Georganopoulos05,Tavecchio08}. For example,
to reproduce the spectra of TeV blazars, \citet{Georganopoulos03a}
have presented a jet model characterized by the deceleration of bulk
velocity along the jet axis. Their model is also used to explain the
X-ray phenomena in kiloparsec jets of powerful radio galaxies and
quasars \citep{Georganopoulos03b}. \citet{Giroletti04a} have
proposed a transverse velocity structure model, described by a
slower external flow surrounding a faster spine, to explain a limb
brightening morphology in Mrk 501 observed by VLBI.
\citet{Ghiselllini08} have used a needle/jet model, which harbors
small active regions (needls) inside a large jet, to account for the
2-3 minutes fast variability of PKS 2155-304. Recent observation of
TeV blazars raise new questions regarding the one-zone uniform jet
model. Such as, the unreally high Lorentz factor (about 50) demanded
by avoiding $\gamma$-$\gamma$ absorption for TeV blazars
\citep{Krawczynski02,Konopelko03,Henri06,Begelman08,Finke08} can not
match the much slower speeds required by Very Long Baseline
Interferometry (VLBI) observations \citep{piner04, piner08}.
Therefore, it is argued that the jet should have velocity structures
to mediate the inconsistence of bulk motions obtained by different
wavebands \citep{Chiaberge00,Trussoni03}.

For unification of BL Lac objects and FR I radio galaxies, the
velocity structures have also been suggested by \citet{Chiaberge00}.
By analyzing the core emissions of FR I radio galaxies observed by
Hubble Space Telescope, they have found that the observed fluxes of
FR I nucleus are over-luminous by a factor of $10-10^4$ than ones
predicted by simple one-zone model, in radio and optical bands.
Therefore they proposed a jet velocity structure, which a fast spine
is surrounded by a slow layer, to reconcile the unification scheme.

In this paper we present the SSC and EC models of AGN jets with
continually longitudinal and transverse bulk velocity structures.
Our study is not completely new, but we include new ingredients and
results. Firstly, the observed apparent speed is obtained by flux
convolution over the whole emitting region with bulk velocity
structures. Secondly, each part of the jet contributes its spectrum
to the observed SED, some important characteristics are ignored in
simple models with two-zone velocity structures compared to the
model with continual velocity structures. Thirdly, some parameters
of jet velocity structures can be constrained by the jet power or
apparent speed , our models are more realistic to discuss the
emissive properties of the jet with velocity structures than
previous models. Fourthly, the jet velocity structures can
reasonably solve the problem of the unification of BL Lacs and FR I
faced by all one-zone models. Fifthly, we find that the effects of
velocity structures upon the observed synchrotron and SSC spectra
strongly depend on the viewing angles. Sixthly, we find that the
longitudinal velocity structures, as the volume elements are
compressed or expanded, have great influences upon the observed
spectra. Finally, the observed EC spectra are weakly affected by
velocity structures compared to the observed synchrotron or SSC
ones.

In Section 2 we present the velocity structures of the jet for the
models calculation and we place these modifications in the context
of emission mechanisms in Section 3. We show the model application
to two BL Lac objetcs (Mrk 421 and 0716+714) controlled by the same
jet power in Section 4, focusing on the unification of BLO with FRI
sources. We then discuss the apparent speeds that govern the
velocity structures of the jet to constrain the observed spectra in
Section 5. In Section 6, we employ the SSC and EC models to one
quasar (3c 279), and discuss the effects of the velocity structures
upon the EC spectra. We finish with discussions and conclusions in
Section 7.

\section{Velocity Structures}

We only select a part of the stationary jet to study. For
simplicity, we adopt the following assumptions: 1. The length of the
studied region is a small scale (i.e., about $1 \times10^ {17 } $
cm) compared with the whole jet. Although the whole jet may be
conical or other morphologies, a cylinder jet is a good
approximation within a short length scale. 2. In the studied region,
the jet is assumed to have the bulk velocity structures in the
longitudinal or transverse directions to the jet axis. The velocity
structures are not resolved by the VLBI observation. 3. We assume
that the jet is stationary and the relativistic electrons are
injected continuously with a power-law energy spectra
$N(\gamma,z_0)=N_0 \gamma^{-s}exp(-\frac{\gamma}{\gamma_{max}})$
\citep{Finke08}. Here we mainly pay attention to the influence of
the velocity structures on radiative spectra, we ignore the
radiative energy loss of the electrons in the studied jet scale.
However, for the jet with the longitudinal velocity structures, as
its volume elements are compressed or expanded, the electron energy
distribution and magnetic field in the jet will be changed follow as
\citet{Georganopoulos04}: $\gamma (z)=\gamma_0 (z/z_0)^{-\xi/3} $,
$N(\gamma,z)=N(\gamma,z_0)(z/z_0)^{-\xi(s+2)/3}$ and
$B(z)=B_0(z/z_0)^{-\xi}$, where the $\xi$ is the index of
longitudinal velocity structures (see the next part), the subscript
"0" denote the values at the studied jet base. 4. For transverse
velocity structures, the magnetic fields are assumed to be a
isotropic distribution in the comoving frame.

\subsection{Longitudinal Velocity Structures}
For a jet with longitudinal velocity structures, we take the Lorentz
factor along the jet to be $\Gamma(z)=\Gamma_0(\frac{z}{z_0})^\xi$,
where $\Gamma_0$ is the Lorentz factor at the base, the $\xi$ is the
index of longitudinal velocity structures. This power-law form was
earlier given by \citet{Marscher80}, subsequently adopted by
\citet{Ghisellini85,Georganopoulos98} and \citet{Li04}. $\xi=0$,
$\xi > 0$, and $\xi < 0$ correspond to transition, acceleration, and
deceleration phases respectively.

VLBI measurements of apparent motion for the parsec-scale radio
knots have often been employed to constrain a combination of Lorentz
factor and viewing angles \citep{Vermeulen94}. Although VLBI
monitoring of the radio knots in blazar jets has revealed several
sources containing knots with apparent speeds extending out to 30c
\citep{Piner07}, the typical values for TeV blazars  are found to be
much more modest, under $5c$ (e.g. \citet{Giroletti04b,piner04}).
\citet{Gopal-Krishna04} have showed that the slow apparent speeds of
the knots of blazars observed by VLBI can be reconciled with the
extremely relativistic bulk motion inferred from TeV flux variations
\citep{Krawczynski02}, if one considers a modest full opening angle
for the parsec-scale jets. They also have showed that the actual
viewing angles, $\theta$, of such conical jets from the
line-of-sight can be substantially larger than those commonly
inferred by VLBI proper motion data \citep{Gopal-Krishna06}.
Recently, they have evaluated the role of the jet opening angle on
certain key parameters (i.e. the viewing angle, the apparent speed
and Doppler factor) that are inferred from VLBI radio observations
of blazar nuclear jets \citep{Gopal-Krishna07}. In these papers,
they have argued that the Doppler boosting of an ultrarelativistic
jet, as well as the apparent proper motions, can greatly vary across
the jet¡¯s cross-section, and then it is important to carry out an
integration of various quantities across the jet cross-section.
Therefore, they have performed an integration of the (boosted)
flux-weighted apparent velocity over the jet cross section to obtain
the weighted observed value of the apparent velocity of the jet,
when the width of a knot cannot be resolved by VLBI observations.

For longitudinal velocity structures the observed apparent speed
($\beta_{app}=\frac{\beta sin\theta}{1-\beta cos\theta}$) is decided
by \citep{Gopal-Krishna04,Gopal-Krishna06,Gopal-Krishna07}

\begin{equation}
\beta_{app,obs} = \frac {\int \beta(z) \delta^2 (z)
dF^{'}_{sync}(z)} {F_{sync}},\label{beta_l}
\end{equation}
where the $F_{sync}$ is the observed flux, and the $
dF^{'}_{sync}(z) $ is the comoving flux in z. In this equation, we
take the radio spectral index to be 0.

Through the radio observation, \citet{Giroletti04b} have found that
the mean value of $\Gamma$ is 3 in the TeV sources. \citet{piner04}
have presented that the jets of TeV blazars have only mildly
relativistic motion which is less than $5c$ on parsec-scales.
\citet{Piner07} have found that the apparent speed distribution
shows a peak at low speed for BL Lac objects. We then take the speed
of $\beta_{app,obs}=3.6c$ to limit $\Gamma_0$.

\subsection{Transverse Velocity Structures}
For a jet with transverse velocity structures, we suppose the jet
bulk velocity to be $\Gamma(R)=\Gamma_c - (\Gamma_c -
\Gamma_m)(R/R_0)^\zeta$, where $\Gamma_c$ and $\Gamma_m$ are the
Lorentz factor of the central and marginal region respectively. In
this work, we only adopt $\zeta$ to be 0 and 1, corresponding to
different transverse velocity structures.

We take the apparent speed given by $\beta_{app,obs}=3.6c$ to
restrict $\Gamma_c$ and $\Gamma_m$. We also use the jet power given
by \citep{Celotti97,Ghiselllini05} $P_{jet}=\int^{R_0}_0 \Gamma^2(R)
U \beta(R) c 2\pi R dR$ to limit $\Gamma_c$ and $\Gamma_m$, where
$U=U_B+U_e+U_p$ is the total energy density in the jet frame.

\section{Radiation Mechanisms}
In this section, the radiation models of continually longitudinal
and transverse velocity structures are constructed respectively.
\subsection{Radiation Model of Longitudinal Velocity Structures }

We present a modified homogeneous SSC model to reproduce the entire
spectral energy distribution. In the following, a prime accent
denotes a parameter expressed in the comoving frame.

For an isotropic electron distribution, the synchrotron emission
coefficient is given by
\begin{equation}
j^{'}_s(\nu^{'}_s,z)
=\frac{1}{4\pi}\int{N_e(\gamma,z)P_e(\nu^{'}_s,\gamma)}d\gamma,
\end{equation}
where $\nu^{'}_s$ is the frequency in the comoving frame, and
$P_e(\nu^{'}_s,\gamma)$ is the mean emission coefficient for a
single electron averaged over an isotropic distribution of pitch
angles. The differential synchrotron luminosity between $z$ and
$z+dz$ is then given by
\begin{equation}
dL^{'}_s(\nu^{'}_s,z,\theta) = 4\pi^2 R_0^2 j^{'}_s(\nu^{'}_s,z) e^{
- \frac{\alpha^{'}_s(\nu^{'}_s,z) R_0}{sin \theta}} dz,\label{dL_s}
\end{equation}
where $R_0$ is the radius of jet, $\alpha^{'}_s(\nu^{'}_s)$ is the
absorption coefficient \citep{Rybicki79} as follows
\begin{equation}
\alpha^{'}_s(\nu^{'}_s,z) = -\frac{1}{8\pi m_e
\nu^{'2}_s}\int\gamma^2\frac{d}{d\gamma}[\frac{N_e(\gamma,z)}{\gamma^2}]P_e(\nu^{'}_s,\gamma)d
\gamma.
\end{equation}
The Doppler factor is given by $ \delta(z,\theta) = \frac
{1}{\Gamma(z)[1-\beta(z)cos\theta]}$ and
$\Gamma(z)=\Gamma_0(\frac{z}{z_0})^\xi$, where
$\Gamma(z)=[1-\beta^2(z)]^{-1/2}$ is the bulk Lorentz factor and
$\beta(z)$ is the velocity in the unit of c. Then the synchrotron
flux density in the observe frame is simply given by
\begin{equation}
 dF_{sync} (\nu_s, z, \theta) =   \frac {1}{4\pi {d_l}^2}
\delta^{(2+\alpha)}(z,\theta)[1+z_{rs}]dL^{'}_s(\nu^{'}_s,z,\theta),
\end{equation}
where $\alpha$ is the spectral index and is taken as 1.0. $d_l$ is
the luminosity distance and $ z_{rs}$ is the redshift. We transfer
$\nu^{'}_s$ into the observer's frame following as:
\begin{equation}
\nu_s(z,\theta) = \frac{\delta(z,\theta)}{[1+z_{rs}]} \nu^{'}_s.
\end{equation}
Finally, we get the total flux density through the integration of
$\nu_s(z,\theta)$ by z.

Calculating the differential SSC emission between $z$ and $z+dz$, we
need to compute soft photon density produced at layers with
different Lorentz factors, which is given by $ u^{'}_{ph} \simeq
u^*_{ph} {\Gamma^2_{rela}} $ \citep{Dermer95, Georganopoulos03a}.
$\Gamma_{rela}$ is the relative Lorentz factor between soft photons
and IC electrons. $u^*_{ph}$ is soft photon energy density in the
local frame. The photon energy density $u^{'}(\nu^{'}_s, z,\theta)$
is given by a sum of a local contribution \citep{Kataoka99}:
\begin{equation}
u^{'}_{loc}(\nu^{'}_s,z)= \frac{3}{4}\frac {4\pi
j_s^{'}(\nu^{'}_s,z)}{c \alpha_s^{'}(\nu^{'}_s,z)}
[1-exp(-\alpha_s^{'}(\nu^{'}_s,z)R_0)],
\end{equation}
and an external contribution:
\begin{eqnarray}
{ u^{'}_{ext}(\nu^{'}_s, z, \theta)=\frac{j_s^{'}(\nu^{'}_s,z)\pi
R_0^2}{c}[ \int^{z-R_0}_{z_0}\Gamma^2_{rela}(z_{s},z,\theta) }
\nonumber\\
{}\frac{exp[-\alpha_s^{'}(\nu^{'}_s,z)(z-z_{s})]dz_{s}
 } {
(z-z_{s})^2} +
\nonumber\\
 {}\int^{z_{max}}_{z+R_0}\Gamma^2_{rela}(z_{s},z,\theta)
\frac{exp[-\alpha_s^{'}(\nu^{'}_s,z)(z_{s}-z)]dz_{s}
 } {
(z_{s}-z)^2}].
\end{eqnarray}
Then the emission coefficient is given by \citep{Inoue96, Kataoka99,
Katarzynski01}
\begin{equation}
j^{'}_c(\nu^{'}_c,z,\theta) = \frac{h}{4\pi} \epsilon^{'}_c
q(\epsilon^{'}_c,z,\theta),
\end{equation}
where $q(\epsilon^{'}_c,z,\theta)$ is the production rate of the
differential photon:
\begin{equation}
q(\epsilon^{'}_c,z,\theta) = \int d \epsilon_s^{'}
n_{ph}(\epsilon_s^{'},z,\theta) \int d \gamma N(\gamma,z)
C(\epsilon_c^{'}, \gamma, \epsilon_s^{'}). \label{q_e}
\end{equation}
$C(\epsilon_c^{'}, \gamma, \epsilon_s^{'})$ is the Compton kernel
given by \citet{Jones68}
\begin{eqnarray}
{ C(\epsilon_c^{'}, \gamma, \epsilon_s^{'}) = \frac{2\pi r^2_e
c}{\gamma^2 \epsilon^{'}_s}[2\kappa ln(\kappa) +
(1+2\kappa)(1-\kappa)+ }
\nonumber\\
{}\frac{(4\epsilon^{'}_s \gamma \kappa)^2}{2(1+4\epsilon^{'}_s
\gamma \kappa)}(1-\kappa)],
\end{eqnarray}
where
\begin{equation} \kappa = \frac{\epsilon^{'}_c}
{4\epsilon^{'}_s\gamma (\gamma-\epsilon^{'}_c) },
\end{equation}
and $r_e$ is the classical electron radius. The integration of
(\ref{q_e}) follows the condition as:
\begin{equation}
\epsilon_s^{'} \leq \epsilon_c^{'} \leq \gamma \frac {4
\epsilon_s^{'} \gamma} {1+4 \epsilon_s^{'} \gamma}.
\end{equation}
Then, we can integrate $dL^{'}_c(\nu^{'}_c,z,\theta)$ to obtain the
total flux density.

In calculating the differential EC emission between $z$ and $z+dz$,
we need to know soft photon energy density produced by external
radiation field, we mainly consider the photons reprocessed by Broad
Line Region (BLR)\citep{Sikora94}.  The photons are assumed to be
distributed as a blackbody with temperature $T_{ext}$. In the
comoving frame of each layer the mean soft photon energy is
blueshifted to $\epsilon^{'}_{ext} \sim \Gamma(z) \epsilon_{ext}
\sim \Gamma(z) k_{B} T_{ext}$, where the $k_B$ is the Boltzmann
constant. The photon energy density is given by $ u^{'}_{ph,ext}
\sim u_{ph,ext} {\Gamma^2(z)}$.

\subsection{Radiation Model of Transverse Velocity Structures }
To the transverse velocity structures, we replace the equation
(\ref{dL_s}) by
\begin{equation} dL^{'}_s(\nu^{'}_s,R, \theta) =
8\pi^2 R j^{'}_s(\nu^{'}_s) (z_{max}-z_0)  e^{ -
\frac{\alpha^{'}_s(\nu^{'}_s) R_0}{sin \theta}} dR.
\end{equation}
Soft photon energy density is given by:
\begin{eqnarray}
u^{'}(\nu^{'}_s, R,\theta)=2 \frac{2\pi j_s^{'}(\nu^{'}_s)}{c}
\int^{\frac{z_{max}-z_0} {2}}_{0}dz  \nonumber\\
\int^{R_0}_{0}dR_{s}\Gamma^2_{rela}(R_{s},R,\theta)
\frac{R_{s}exp[-\alpha_s^{'}(\nu^{'}_s)(R^{2}+z^2)^{1/2}] } {
(R^2+z^2)},
\end{eqnarray}
where $R_s$ is the radius of producing the soft photons. Since the
studied length is assumed to be short compared to the whole jet, we
take the soft photon density at $\frac{z_{max}-z_0} {2}$ in
different z as an approximation . We find that this simplicity is
reasonable in the following calculation.

\section{Velocity Structure Constrained by Jet Power}

In this section, we apply the transverse velocity structures
constrained by jet power to the SEDs of Mrk 421 and 0716+714, and
discuss their properties in the unification of BL Lac objects and FR
I radio galaxies.

\begin{figure}
\begin{center}
 \FigureFile(80mm,80mm){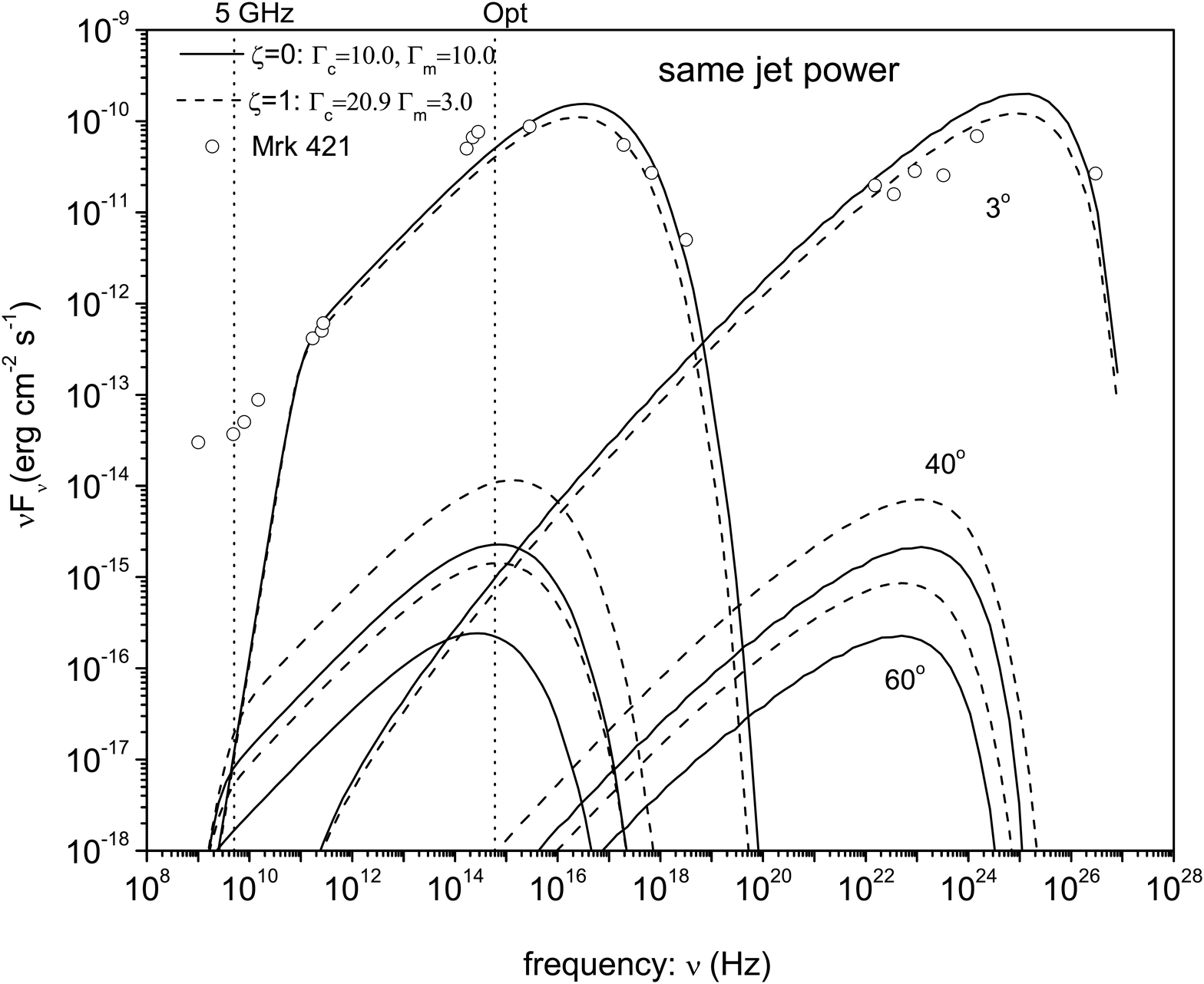}
  \end{center}
  \caption{The approximately
fitting for data of Mrk 421 (from \citet{Macomb96} and NED) and the
extrapolated SEDs of parent population. The solid lines represent
the emission of jet with uniform velocity structures ($\zeta=0$) and
the dash lines denote the jet with velocity structures ($\zeta=1$).
The upper two curves report the emission at angles of $3^{\circ}$,
the mid and lower two curves are the cases of parent population of
BL Lacs extrapolating the viewing angles to $40^{\circ}$ and
$60^{\circ}$. \label{fig.1}}
\end{figure}

We select the well observed Mrk 421 that is a typically high energy
peaked BL Lac (HBL). The SED of Mrk 421 is obtained by fitting the
observed core data (for the quiescent state of Mrk 421, the data is
from the \citet{Macomb96} and NED) with our model. The SEDs of its
parent population are then given by increasing the viewing angles.
In this work, we adopt two kinds of transverse velocity structures
$\zeta=0$ (uniform velocity structures) and $\zeta=1$ (bulk Lorentz
factor linearly lessen from the center to the margin). The observed
data of Mrk 421 are fitted using the parameters listed in Table 1.
Its corresponding SEDs in different viewing angles are calculated to
show the implications of velocity structures. \citet{Konopelko03}
have used the one-zone SSC model to fit the lower state SEDs of Mrk
421 and obtained the parameters of $s=1.75$,
$\gamma_{max}=3\times10^5$, $ B=0.10G$; \citet{Blazejowski05} have
presented the parameters of $\delta= 10$, $B =0.405G$, $R=
0.7\times10^{16} cm$. The parameters in our model are comparable to
their ones. We set $\Gamma_c=\Gamma_m=10$ for the jet with uniform
structure and $\Gamma_m=3$ for the jet with velocity structure
(\citet{Ghiselllini05}; \citet{Giroletti04a}; \citet{Tavecchio08}).
Assuming the same jet power for two structures, we can obtain
$\Gamma_c $ for the jet with velocity structure. The SEDs of Mrk 421
are shown in Fig.~\ref{fig.1}. The SEDs of two velocity structures
show very small difference in the small viewing angle ($3^{\circ}$).
Under the viewing angle of parent population (FR I) ($60^{\circ}$ ),
the SEDs have large difference, in which the flux with velocity
structure is much larger than ones with uniform structure. The
velocity structure can resolve the problem that the observed FR I
nuclei are over-luminous by a factor of $10-10^4$ than ones
predicted by simple one-zone model in optical and radio bands. In
order to compare the observed and predicted fluxes of parent
population, we also present two vertical lines denoting the optical
(V band) and 5 GHz radio band respectively in Fig.1. It is noted
that the radio fluxes given by the model are extrapolated from the
infrared fluxes.

\begin{figure}
\begin{center}
 \FigureFile(80mm,80mm){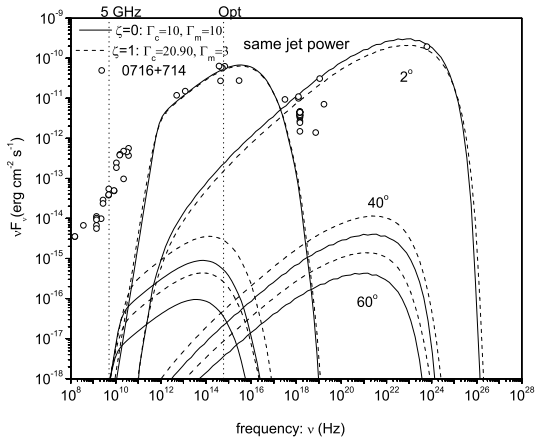}
  \end{center}
  \caption{The approximately
fitting for NED data of 0716+714 and the extrapolated SEDs of parent
population. The solid lines represent the jet with uniform velocity
structures ($\zeta=0$) and the dash lines denote the jet with
velocity structures ($\zeta=1$). The upper two curves report the
emission at angles of $2^{\circ}$, the mid and lower two  curves are
the cases of parent population of BL Lacs extrapolating the viewing
angles to $40^{\circ}$ and $60^{\circ}$.\label{fig.2}}
\end{figure}

0716+714 is a typical low energy peaked BL Lac (LBL). We also list
the parameters in the Table 1. \citet{Giommi99} gave the magnetic
field to be $B>0.9 $G. \citet{Tagliaferri03} got the parameters of
$B=2.5$G, $R=2\times10^{16}$, $\theta=3.4^{\circ}$. These parameters
are consistent with ones given by our model. The SEDs of 0716+714
given by the model are shown in Fig.~\ref{fig.2} . It is shown that
the SEDs have obvious difference for two velocity structures under
large viewing angles ($40^{\circ}$ or $60^{\circ}$).

\begin{figure}
\begin{center}
 \FigureFile(80mm,80mm){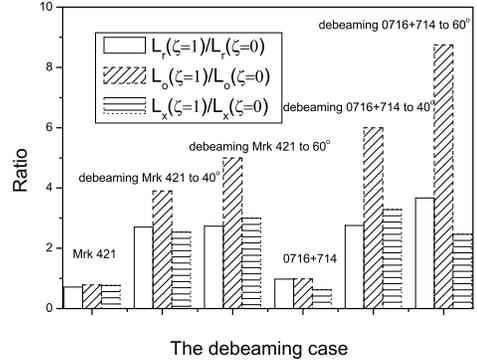}
  \end{center}
  \caption{The ratio of luminosity
between $\zeta=1$ and $\zeta=0$ under different band and viewing
angles.\label{fig.3}}
\end{figure}

\begin{figure}
\begin{center}
 \FigureFile(80mm,80mm){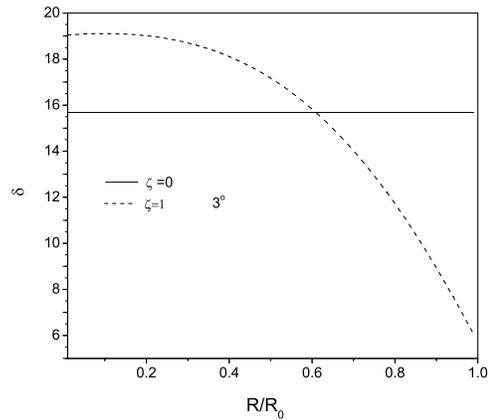}
  \end{center}
  \caption{Under
$\theta=3^{\circ}$, the Doppler factors vary with R for different
velocity structures. Solid line denotes the case of uniform velocity
structures and the dash line is in velocity
structures.\label{fig.4}}
\end{figure}

\begin{figure}
\begin{center}
 \FigureFile(80mm,80mm){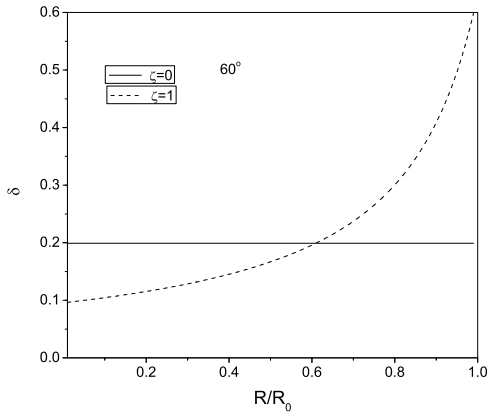}
  \end{center}
  \caption{Under
$\theta=60^{\circ}$, the Doppler factors vary with R for different
velocity structures. Solid line denotes the case of uniform velocity
structures and the dash line shows the case of velocity
structures.\label{fig.5}}
\end{figure}

In the Fig.~\ref{fig.3}, we give the luminosity ratio between the
jets with $\zeta=1$ and $\zeta=0$ in three bands: radio (9GHz),
optical (V band), and X-rays (1keV) for Mrk 421, 0716+714, and their
parent population. It is shown that the ratio can reach $2-9$ in
three bands under large viewing angles corresponding to parent
population. Understanding these properties, we present the
variations of Doppler factors along the jet radius R for different
velocity structures under the viewing angles of $3^{\circ}$ and
$60^{\circ}$, shown in Fig.~\ref{fig.4} and \ref{fig.5}. Under
$3^{\circ}$ (see the Fig.~\ref{fig.4}), the Doppler factor of the
jet with velocity structure rapidly decreases with R and is smaller
than ones of the jet with uniform velocity structure when $R > 0.6
R_0 $. However, the increase of the Doppler factor with R appears
under $60^{\circ}$ (Fig.~\ref{fig.5}).

\begin{figure}
\begin{center}
 \FigureFile(80mm,80mm){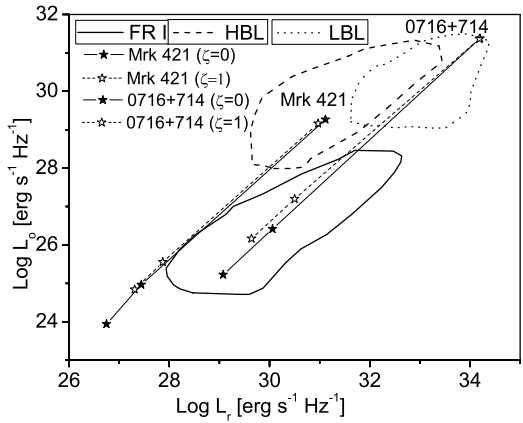}
  \end{center}
  \caption{Debeaming trails with
the models of continual velocity structures in the radio-optical
luminosity plane for Mrk 421 and 0716+714 (the solid lines denote
$\zeta=0$ and dash ones are $\zeta=1$), where upper pentagrams
correspond to the luminosity of BL Lacs at $\theta=2^{\circ}$ or $
3^{\circ}$ and lower ones are the predicted luminosity of parent
population at $\theta=40^{\circ}$ and $60^{\circ}$. The three
regions enclosed by curves respectively describe the range for three
samples, i.e., HBL (dash curves), LBL (dot curves), and FR I (solid
curves) in realistic observations \citep{Chiaberge00}.
\label{fig.6}}
\end{figure}

In the unification scheme, it is believed that the appearance of
AGNs strongly depends on the viewing angles and obscuration instead
of intrinsic physical properties. For low luminosity radio-loud
objects, such as BL Lac objects and FR I radio galaxies, the
relativistic beaming effects caused by viewing angles should play an
important role in observation. However, \citet{Chiaberge00} have
found that the observed FR I nuclei are over-luminous by a factor of
$10-10^4$ than ones predicted by simple one-zone model in the radio
and optical bands. They have argued that a radial velocity
structure, which a fast spine surrounded by a slow layer, can
reconcile the above contradiction. We apply the velocity structures
described previously to the unification of BLO and FRI. In
Fig.~\ref{fig.6}, we show the debeaming trails with different
velocity structures in the radio-optical luminosity plane for Mrk
421 and 0716+714. For Mrk 421, when the viewing angle is increased
to $60^{\circ}$, the predicted optical and radio luminosity nearly
come into the region of FRI in case of velocity structure, however
they are less-luminous in case of uniform structure. For 0716+714,
the predicted luminosity fully fall into the region of FRI in
different velocity structures.

\begin{figure}
\begin{center}
 \FigureFile(80mm,80mm){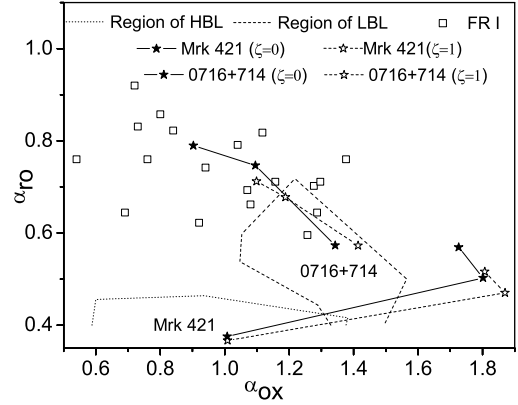}
  \end{center}
  \caption{Debeaming trails in the
$\alpha_{ox}$ - $\alpha_{ro}$ plane for Mrk 421 and 0716+714. The
solid denote $\zeta=0$ and dash lines is $\zeta=1$. The lower
pentagrams represent BL Lac, while the upper ones describe the
parent population of BL Lac observed at $\theta = 40^{\circ}$ and
$60^{\circ}$. The data of FR I, HBL, and LBL are given by
\citet{Fossati98,Trussoni03}. \label{fig.7}}
\end{figure}

The $\alpha_{ro}$-$\alpha_{ox}$ planes for Mrk 421, 0716+714 and FRI
are presented in the Fig.~\ref{fig.7}, in which the lines denote the
debeaming trails of Mrk 421 and 0716+714. The plotted data of FRI
and BLO are from the \citet{Fossati98} and \citet{Trussoni03}. For
0716+714, the debeamed indices fall into the region of FRI under
different velocity structures. However, the debeamed indices of
Mrk421 are marginally in the region of FRI. This implies some
intrinsic physical difference exist between HBL and LBL.

\section{Velocity Structure Constrained by Apparent Speed}

In this section, we use the observed apparent speed to constrain the
jet velocity structures and discuss its effect on the SEDs of BL Lac
objects shown in Mrk 421 and 0716+714.

\begin{figure}
\begin{center}
 \FigureFile(80mm,80mm){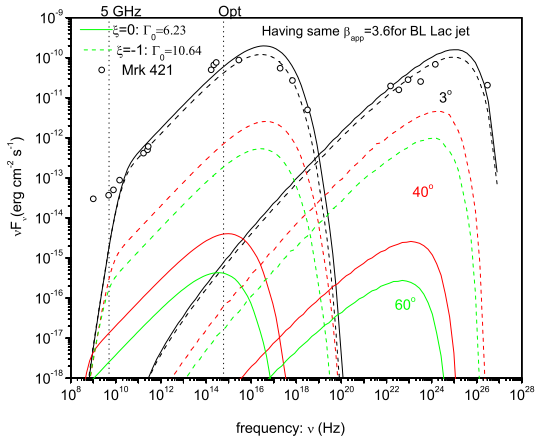}
  \end{center}
  \caption{Under
$\beta_{app}=3.6c$, the approximately fitting for data of Mrk 421
(from \citet{Macomb96} and NED) and the extrapolated SEDs of parent
population (Note that we use different parameters for different
velocity structures, see the Table 1.). The solid lines represent
the emission of jet with uniform velocity structures ($\xi=0$) and
the dash lines denote the jet with velocity structures ($\xi=-1$).
The upper curves show the emission at angles of $3^{\circ}$, the red
and green curves present parent population of BL Lacs with the
viewing angles of $40^{\circ}$ and $60^{\circ}$.\label{fig.8}}
\end{figure}

\begin{figure}
\begin{center}
 \FigureFile(80mm,80mm){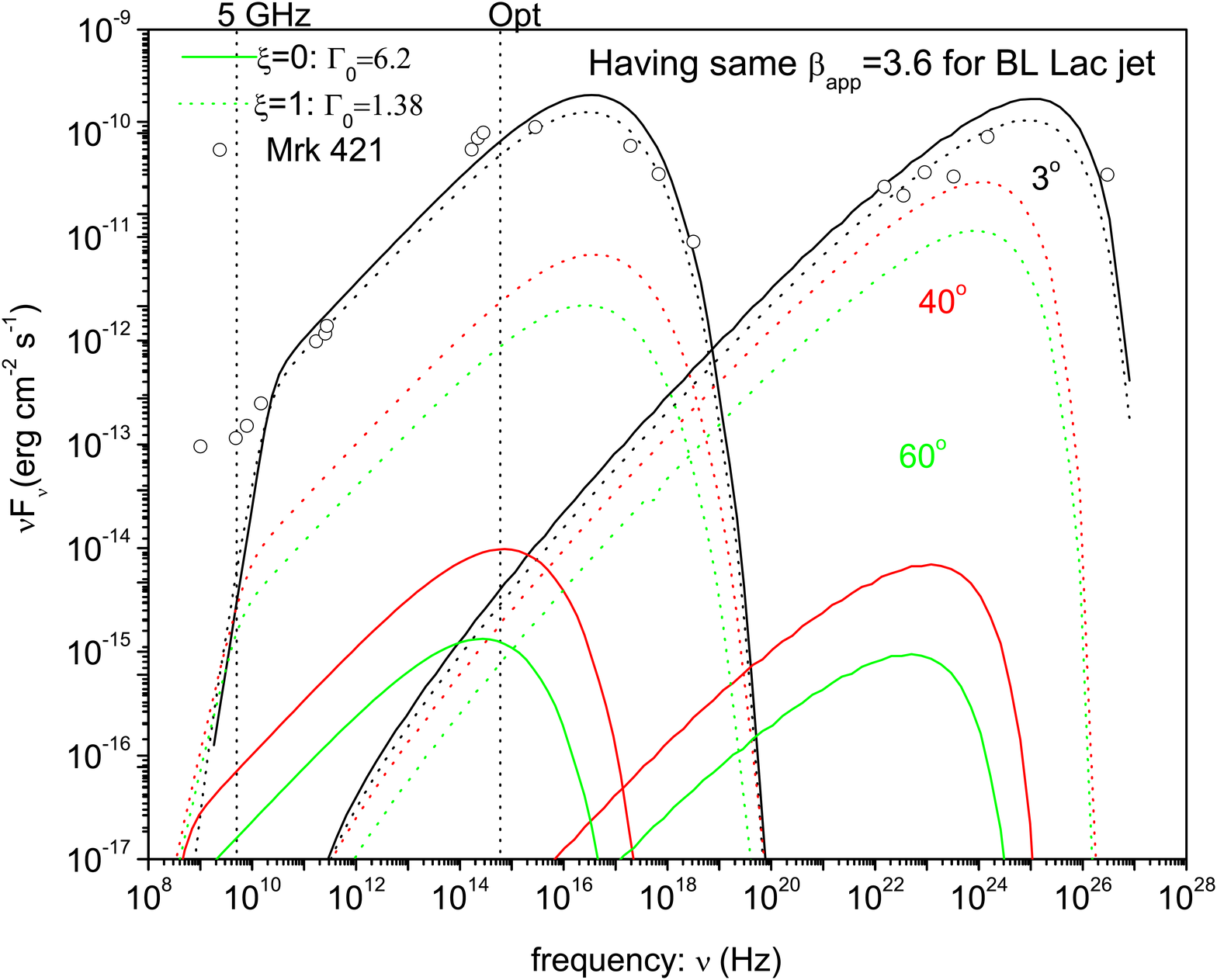}
  \end{center}
  \caption{Under
$\beta_{app}=3.6c$, the approximately fitting for data of Mrk 421
(from \citet{Macomb96} and NED) (Note that we use different
parameters for different velocity structures, see the Table 1.) and
the extrapolated SEDs of parent population. The solid lines
represent the emission of jet with uniform velocity structures
($\xi=0$) and the dot lines denote the jet with velocity structures
($\xi=1$). The upper curves report the emission at angles of
$3^{\circ}$, the red and green curves present parent population of
BL Lacs with the viewing angles of $40^{\circ}$ and
$60^{\circ}$.\label{fig.9}}
\end{figure}

\begin{figure}
\begin{center}
 \FigureFile(80mm,80mm){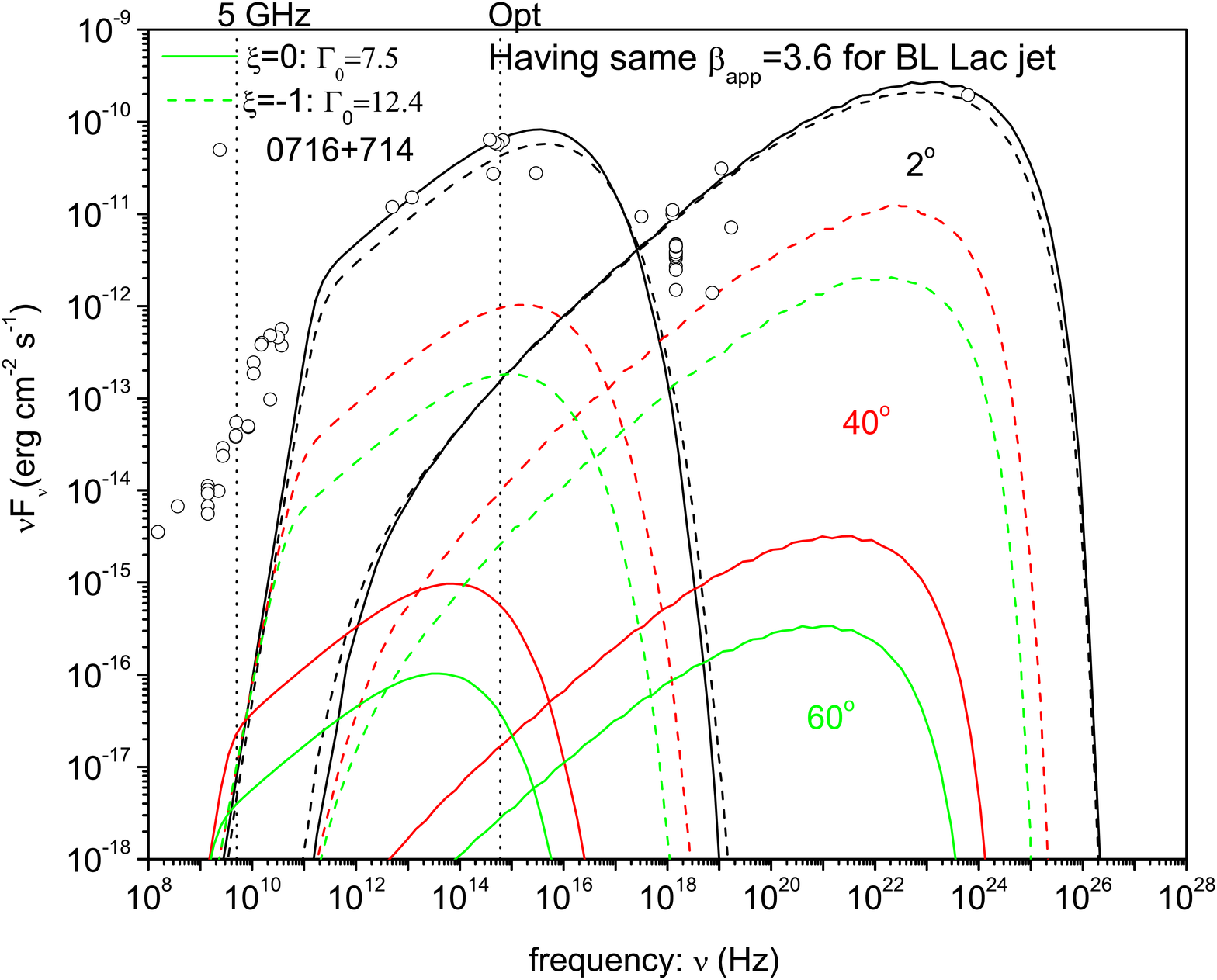}
  \end{center}
  \caption{Under
$\beta_{app}=3.6c$, the approximately fitting for NED data of
0716+714 (Note that we use different parameters for different
velocity structures, see the Table 1.) and the extrapolated SEDs of
parent population. The solid lines represent the jet with uniform
velocity structures ($\xi=0$) and the dash lines denote the jet with
velocity structures ($\xi=-1$). The upper curves report the emission
at angles of $2^{\circ}$, the red and green curves present parent
population of BL Lacs with the viewing angles of $40^{\circ}$ and
$60^{\circ}$.\label{fig.10}}
\end{figure}

\begin{figure}
\begin{center}
 \FigureFile(80mm,80mm){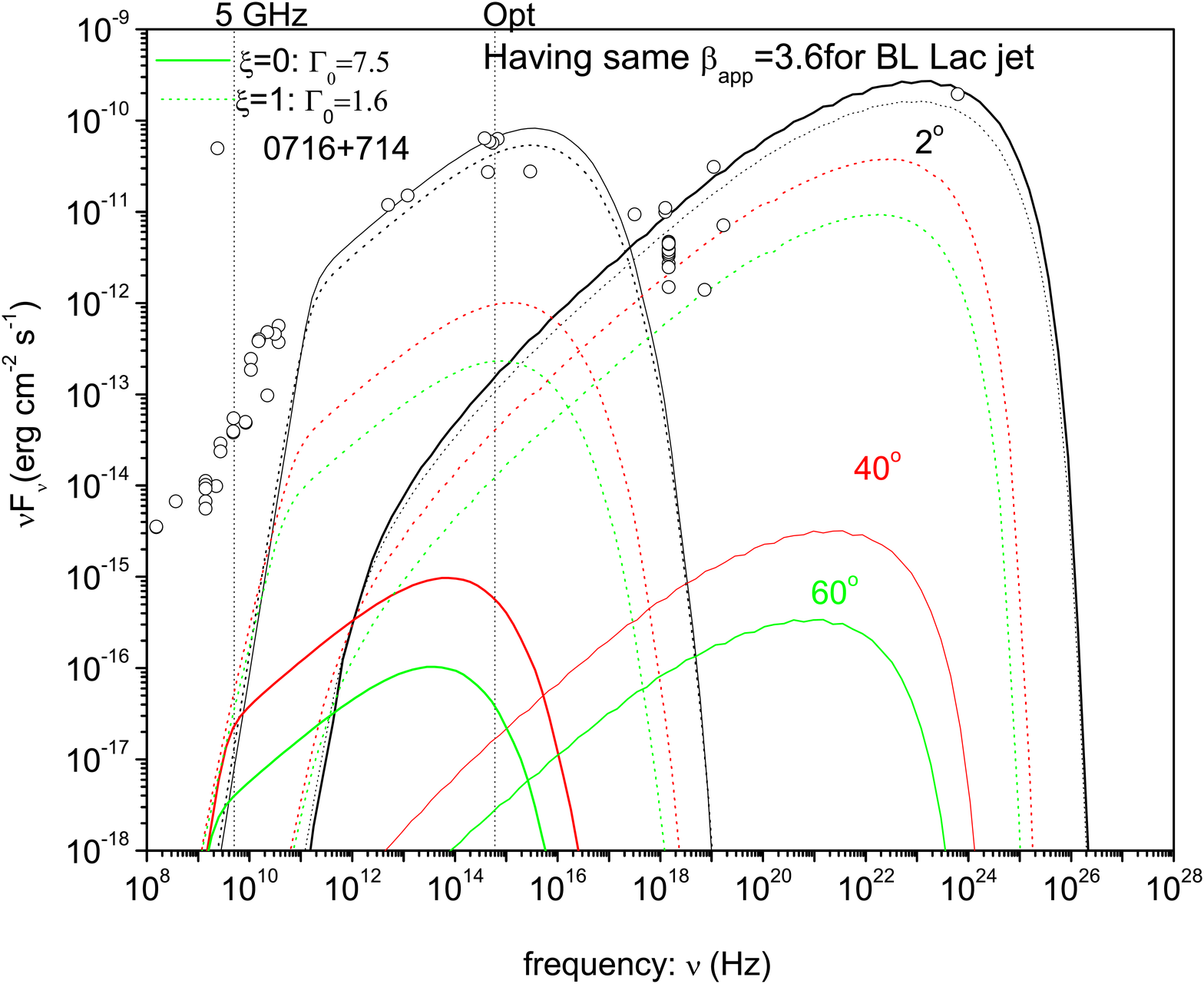}
  \end{center}
  \caption{Under $\beta_{app}=3.6c$, the approximately fitting for
NED data of 0716+714 (Note that we use different parameters for
different velocity structures, see the Table 1.) and the
extrapolated SEDs of parent population. The solid lines represent
the jet with uniform velocity structures ($\xi=0$) and the dot lines
denote the jet with velocity structures ($\xi=1$). The upper curves
report the emission at angles of $2^{\circ}$, the red and green
curves are present parent population of BL Lacs with the viewing
angles of $40^{\circ}$ and $60^{\circ}$.\label{fig.11}}
\end{figure}

\begin{figure}
\begin{center}
 \FigureFile(80mm,80mm){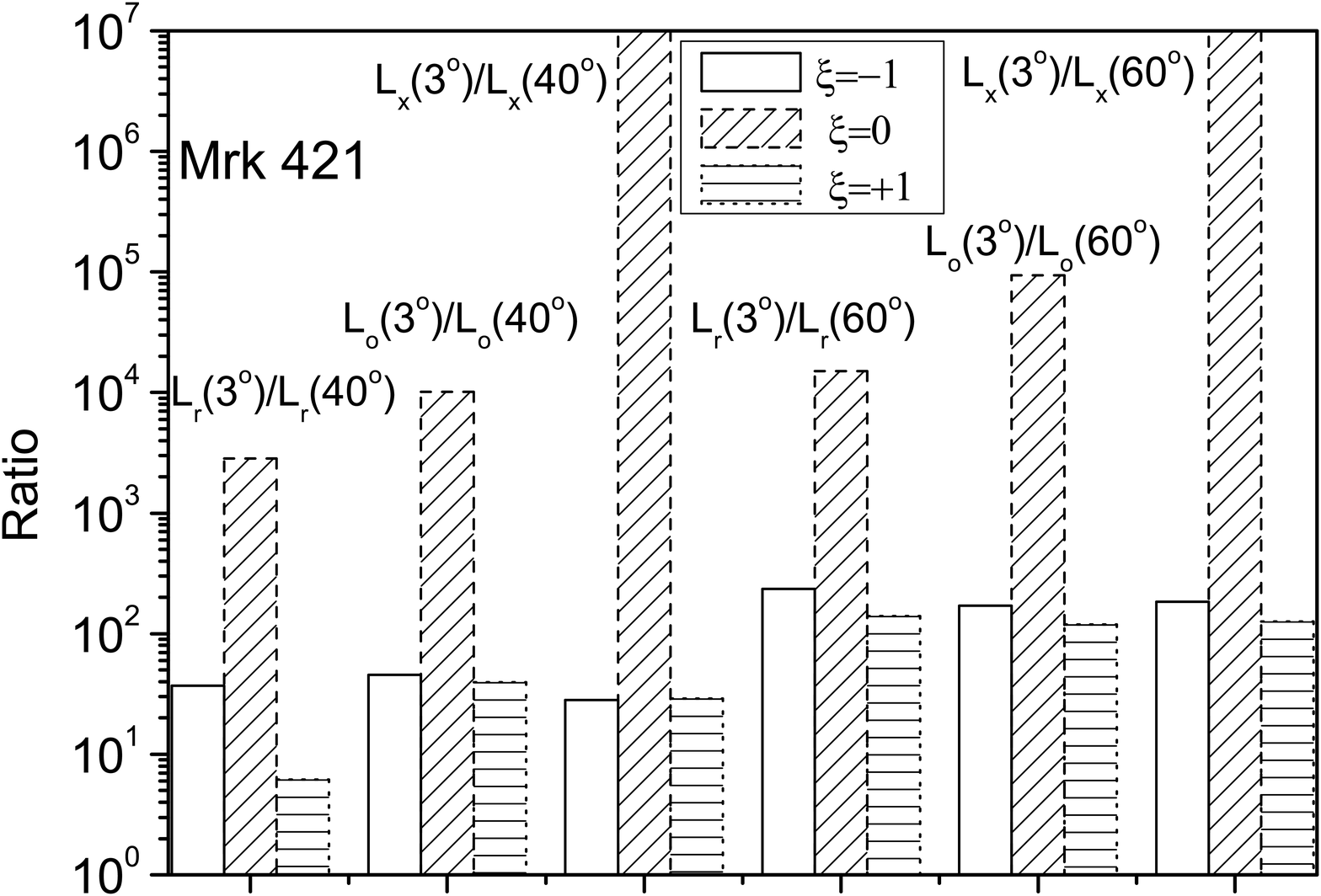}
  \end{center}
  \caption{The ratio of luminosity
between different viewing angles under different $\xi$ and bands for
Mrk 421.\label{fig.12}}
\end{figure}

\begin{figure}
\begin{center}
 \FigureFile(80mm,80mm){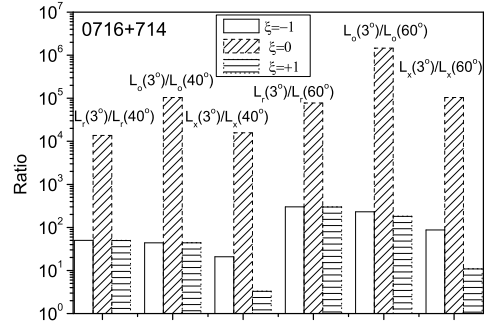}
  \end{center}
  \caption{The ratio of luminosity
between different viewing angles under different $\xi$ and bands for
0716+714. \label{fig.13}}
\end{figure}

\begin{figure}
\begin{center}
 \FigureFile(80mm,80mm){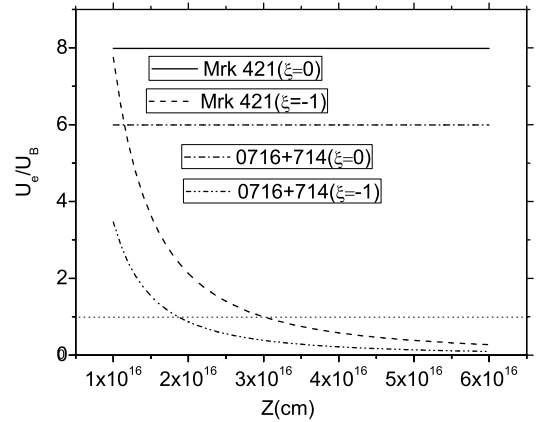}
  \end{center}
\caption{The change of $U_e/U_B$ with z in uniform and deceleration
velocity structures. The dot line is $U_e=U_B$, solid ($\xi=0$) and
dash lines ($\xi=-1$) show Mrk 421, and the dash-dot ($\xi=0$) and
dash-dot-dot lines ($\xi=-1$) denote the 0716+714.\label{fig.14}}
\end{figure}

We take $\beta_{app}=3.6c$ to limit longitudinal and transverse
velocity structures and calculate the SEDs of the jet with velocity
structures to fit the observed data of Mrk 421 and 0716+714. Under
$\beta_{app}=3.6c$, adopting $3^{\circ}$ or $2^{\circ}$ as the
viewing angles we obtain $\Gamma_0$ for three kinds of longitudinal
velocity structures, e.g. uniform ($\xi=0$), decelerating($\xi=-1$),
and accelerating($\xi=1$), which are listed in the figures. The SEDs
of Mrk 421, 0716+714 and their parent populations under different
velocity structures are shown in Fig.~\ref{fig.8}, \ref{fig.9},
\ref{fig.10}, and~\ref{fig.11}. It is shown that the elementary
volume expansion ($\xi>0$) or compression ($\xi<0$) change the
electron energy distribution in longitudinal bulk velocity
structures and cause the parameters fitting the observed data to be
large difference between uniform and velocity structure models (see
the Table 1). In the decelerating velocity structures, we use
smaller parameters of $N_0$, $\gamma_{max}$ and $B$ to fit the
observed SEDs compared to the uniform structure. On the contrary, we
must use larger parameters in the accelerating models to fit the
SEDs. When the velocity structures exist, the flux $\nu F_{\nu}$
decreases slowly along the increase of the viewing angles (such as
$40^{\circ}$ or $60^{\circ}$). However, the flux $\nu F_{\nu}$
decreases quickly in the uniform structure. In the Fig.~\ref{fig.12}
and \ref{fig.13}, we give the luminosity ratio between different
viewing angles in the same band. In the uniform jet, the ratio
changes greatly with the viewing angles. The energy equipartition
between electron and magnetic field could not be
true\citep{Hardcastle98}. We find that for the acceleration velocity
structures $U_e/U_B$ has larger value (about 300) at the outside of
studied jet scale, however, for the deceleration velocity structures
the $U_e$ is larger than $U_B$ at the inner of studied jet scale,
along the jet, $U_B$ is became dominant at the outer (see the
Fig.~\ref{fig.14}).

\begin{figure}
\begin{center}
 \FigureFile(80mm,80mm){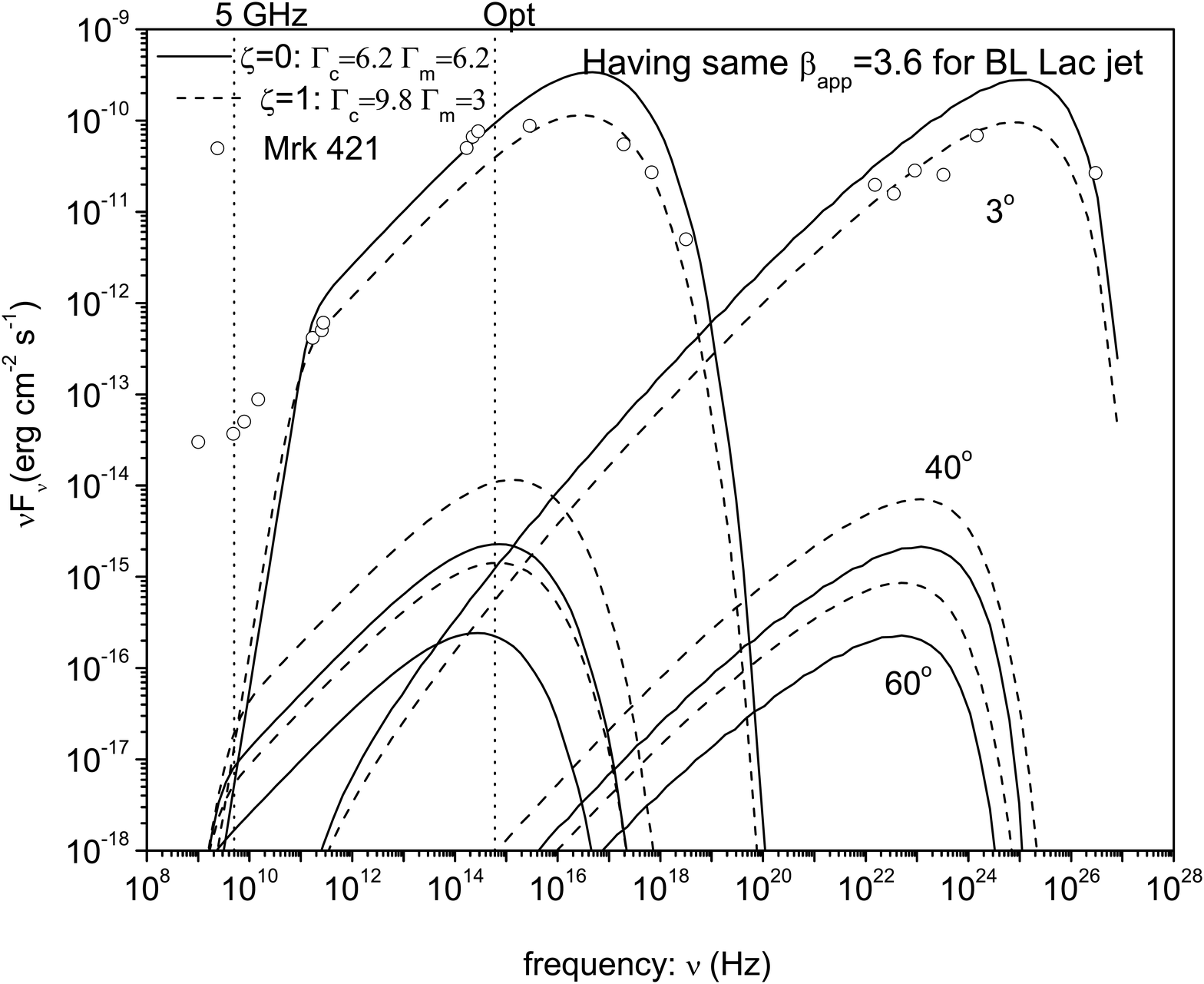}
  \end{center}
  \caption{Under
$\beta_{app}=3.6c$, the approximately fitting for data of Mrk 421
(from \citet{Macomb96} and NED) and the extrapolated SEDs of parent
population. The solid lines represent the emission of jet with
uniform velocity structures ($\zeta=0$) and the dot lines denote the
jet with velocity structures ($\zeta=1$). The upper curves show the
emission at angles of $3^{\circ}$, the mid and lower curves present
parent population of BL Lacs with the viewing angles of $40^{\circ}$
and $60^{\circ}$.\label{fig.15}}
\end{figure}

\begin{figure}
\begin{center}
 \FigureFile(80mm,80mm){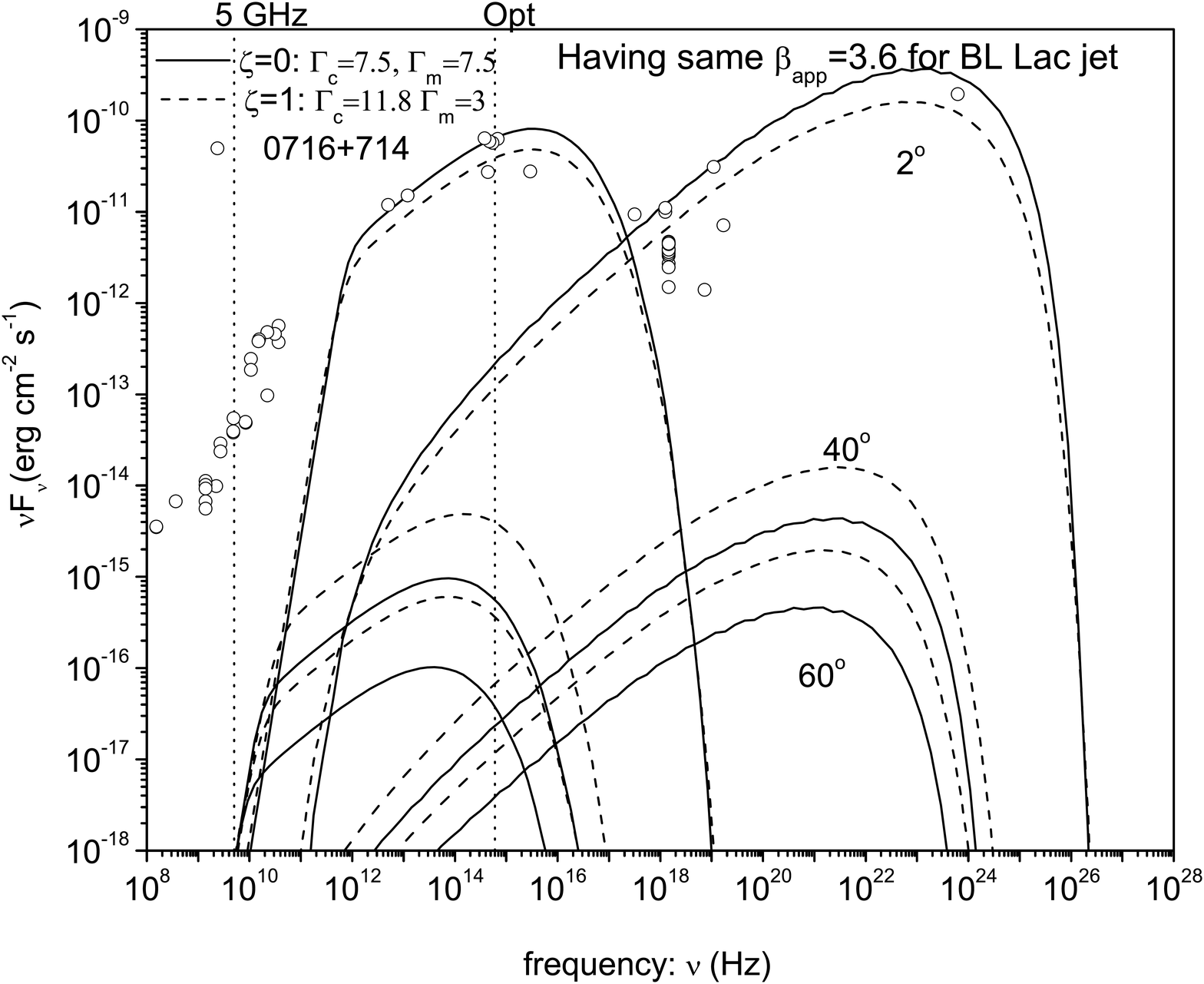}
  \end{center}
  \caption{Under
$\beta_{app}=3.6c$, the approximately fitting for NED data of
0716+714 and the extrapolated SEDs of parent population. The solid
lines represent the jet with uniform velocity structures ($\zeta=0$)
and the dash lines denote the jet with velocity structures
($\zeta=1$). The upper curves show the emission at angles of
$2^{\circ}$, the mid and lower curves present parent population of
BL Lacs with the viewing angles of $40^{\circ}$ and
$60^{\circ}$.\label{fig.16}}
\end{figure}

To the transverse velocity structures (Fig.~\ref{fig.15} and
~\ref{fig.16}), we adopt two kinds of velocity structures, e.g.
$\zeta=0$ (uniform velocity structure) and $\zeta=1$ (bulk Lorentz
factor linearly decreases from the center to the margin). We set
$\Gamma_m=3$ and obtain $\Gamma_c$ restricted by $\beta_{app}=3.6$,
which is listed in corresponding figures. The SEDs of Mrk 421 and
0716+714 in different viewing angles are shown in Fig.~\ref{fig.15}
and ~\ref{fig.16}. We show that the lower luminosity is common to
the jets with uniform velocity structure in large viewing angles,
and that the way of solving the paradox of BLO and FRI unification
is to consider the velocity structure, where all photons are not
produced in same Doppler factor.

\section{EC Spectra with Velocity Structure}

\begin{figure}
\begin{center}
 \FigureFile(80mm,80mm){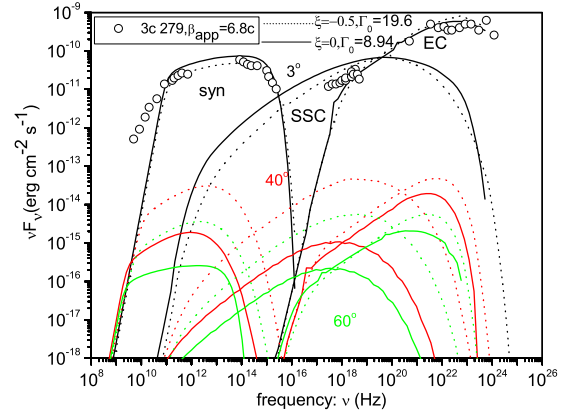}
  \end{center}
  \caption{Under
$\beta_{app}=6.8c$, the fit curve for data of FSRQ 3c 279 from the
\citet{Inoue96} with the models of uniform ($\xi=0$) or velocity
structures ($\xi=-0.5$) (Note that we use different parameters for
different velocity structures, see the Table 1.) and the
extrapolated SED of parent population. The solid lines represent the
jet with uniform velocity structures ($\xi=0$) and the dot lines
denote the jet with velocity structures ($\xi=-0.5$). The upper
curves show the emission at the angle of $3^{\circ}$, the red and
green curves corresponds to the parent population of quasar with the
viewing angles of $40^{\circ}$ and $60^{\circ}$.  \label{fig.17}}
\end{figure}

\citet{Piner07} have proposed that the mean fastest apparent speed
for the quasars is $6.8 \pm  1.1c$, as the previous discussion we
will use this observed apparent speed to constrain the parameters of
velocity structures (i.e., $\Gamma_0=19.6$ for the $\xi=-0.5$;
$\Gamma_0=8.4$ for the $\xi=0$). In the Fig.~\ref{fig.17}, under
$\beta_{app}=6.8c$, we present the fit for the data of FSRQ 3c 279
\citep{Inoue96}. In the model, we set $T_{ext}=10eV$ (near the
energy of hydrogen Lyman-$\alpha$ photons) and $U_{ext}=6.6 \times
10^{-5}$ erg $cm^{-3}$ for external photons, other parameters are
shown in the Table 1. The solid lines represent the jet with uniform
velocity structures ($\xi=0$) and the dot lines denote the jet with
velocity structures ($\xi=-0.5$). The upper curves show the emission
with the angle of $3^{\circ}$, the mid and lower curves correspond
to the parent population of quasars with the viewing angles of
$40^{\circ}$ and $60^{\circ}$. From the figure, firstly, we find
that the observed EC component is weakly affected by velocity
structures compared to synchrotron or SSC ones. The reason is that
for continual jet the luminosity $L_{EC}$ has the relation of
$L_{EC} \sim N_e \delta^3 \dot\gamma^{'}_{EC} \sim N_e \delta^3
U^{'}_{ext} \sim \Gamma^2 \delta^3 U_{ext} N_e $, the integration of
$\Gamma^2 \delta^3$ over z gives less discrepancy between the models
of $\xi=-0.5$ and $\xi=0$.

\section{Discussions and Conclusions}
In this work we mainly pay attention to the influence of the bulk
velocity structures on radiative spectra and ignore the radiative
energy loss of the electrons in the studied jet scale. It is
reasonable if we suppose ongoing particles acceleration in situ can
compensate the radiative loss. However, radiative losses become
important for high energy electrons producing the high energy part
of synchrotron and IC emission, when the particles acceleration
mechanism do not offset the radiative loss. In this situation, the
high energy part of synchrotron and IC emission will be mainly
originated at the base of jet, and their spectra will be steeper
with the distance.

In this work, the synchrotron self-Compton (SSC) models and External
Compton (EC) models of AGN jets with continually longitudinal and
transverse bulk velocity structures are constructed. But our models
can not discriminate a lateral and longitudinal deceleration
scenario. For the former scenario we might observe the limb
brightening in VLBI \citep{Giroletti04a}. For the longitudinal
deceleration scenario it should exist an outward-increasing
radio-to-X-ray ratio \citep{Georganopoulos03b}. But they might be
interrelated, for example, in a faster spine and slower layer model
when the IC power is compared with the total bulk kinetic power. The
Compton rocked effect will cause the spine to recoil and decelerate,
and form the longitudinal velocity structures \citep{Ghiselllini05}.
Similarly, if the gradual entrainment from external medium cause the
jet deceleration, it would produce velocity gradients across the
jet. A faster spine and a slower sheath are formed.

The rapid TeV flares observed for PKS 2155-304 and Mrk 501
\citep{Aharonian07,Albert07} appears to indicate large bulk Lorentz
factors. The systematic differences of the Doppler factors inferred
from different waveband observations indicate that the jet holds
velocity structures within sub-parsec to parsec scale. Based on our
calculation, large bulk Lorentz factor in the base of jet for
longitudinal velocity structures or in the spine for transverse
velocity structures, can satisfy the TeV photon escape, where the
apparent speed is moderate.

In the unified schemes, we find that beside the viewing angles, the
bulk velocity structures play an important role in the Doppler
beaming pattern. Firstly, from the observed data to deduce the
intrinsic quantities, the velocity structures and the viewing angles
should be considered at one time. In fact, if the jets have velocity
structures, their intrinsic flux will be difficultly deduced under
different viewing angles. Secondly, the velocity structures also
bring some statistical discrepancy between observation and theory in
simple one-zone model \citep{Chiaberge00,Henri06}. In one-zone
model, if $\theta \leq 1/ \Gamma$, one has $1\leq\delta\leq2\Gamma$
except $\delta\approx1/\Gamma$. This means that for a few Doppler
amplified sources, one expects a large number of unbeamed
counterparts. In this work, we find that under larger viewing angles
the velocity structures limited by apparent speed have larger effect
on the observed flux. The statistic of the flux limited sources will
provide a framework to explore the velocity structures of AGN jets.

\begin{figure}
\begin{center}
 \FigureFile(80mm,80mm){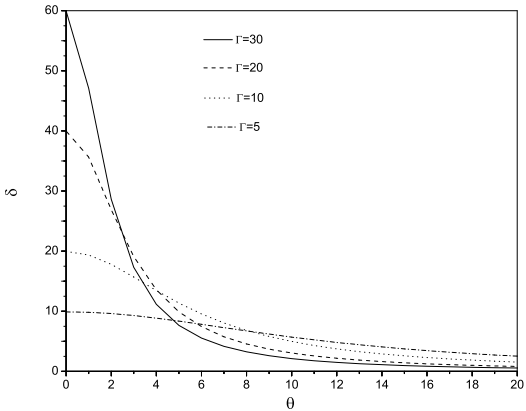}
  \end{center}
  \caption{For different bulk
Lorentz factor of 30, 20, 10 and 5, the Doppler factors vary with
the view angles.\label{fig.18}}
\end{figure}

In the Fig.~\ref{fig.18} also given by \citet{Urry95}, we present
the relations between Doppler factor, Lorentz factor and viewing
angle. The curve lines for different Lorentz factors cross each
other within the viewing angles of $1^{\circ}$ to $10^{\circ}$. They
show that large bulk Lorentz factor does not always produce large
Doppler factor when the viewing angle increases. Furthermore, the
Doppler factor with large Lorentz factor decreases quickly under the
increase of viewing angles. These properties cause complex effects
on emissive spectra.

The observed apparent speed might be the results of flux convolution
over the resolution region, which can solve the inconsistence of
bulk motions obtained by different wavebands. We find that the
influences of the velocity structures upon the observed spectra
heavily depend on the viewing angles. Under the jet with velocity
structure constrained by jet power and apparent speed, we show that
the SEDs have large difference for the jets with and without
velocity structure in large viewing angle. For the jet with
transverse and longitudinal deceleration velocity structure, its
flux is much larger than one with uniform structure. Especially, in
longitudinal bulk velocity structures, as the elementary volume
expansion ($\xi>0$) or compression ($\xi<0$) greatly change the
electron energy distribution, we use smaller values of $N_0$,
$\gamma_{max}$ and $B$ than the uniform ones to fit the SEDs; on the
contrary, we must use larger values for the accelerating models to
get fitting. We find that for the deceleration velocity structures
the $U_e$ is larger than $U_B$ at the inner of studied jet scale,
along the jet, $U_B$ is became dominant at the outer. The EC
observed spectra are weakly affected by velocity structures compared
to synchrotron and SSC ones.

Nowadays, the VLBA can not resolve the velocity structures with
sub-parsec to parsec scale, the AGN's unified schemes exit a large
uncertainty. The TeV emission is easily produced in the jet with
large velocity structures and indicates that TeV blazars should have
the jet bulk velocity structures within sub-parsec to parsec scale.
\\

We thank the anonymous referee for useful comments that led to
improvements in the paper. We acknowledge the financial supports
from the National Natural Science Foundation of China 10673028,
10778702 and 10803018, and the National Basic Research Program of
China (973 Program 2009CB824800). This research made use of the
NASA/IPAC Extragalactic Database (NED) which is operated by the Jet
Propulsion Laboratory, California Institute of Technology, under
contract with the National Aeronautics and Space Administration.

\newpage

\begin{table}[!h]
\tabcolsep 2mm \caption{Input parameters of the models for the Mrk
421, 0716+714 and 3c 279. The unit of the $z_0$, $z_{max}$, and $R$
are cm; B is Gauss}
\begin{center}
\begin{tabular}{r@{ }lr@{   }lr@{ }lr@{  }lr@{ }lr@{ }lr@{ }lr@{}lr@{ }l}
\hline \multicolumn{1}{c}{
Objects}&\multicolumn{1}{c}{$n_0$}&\multicolumn{1}{c}{$\gamma_{max}$}&
\multicolumn{1}{c}{Index}&\multicolumn{1}{c}{$z_0$}&\multicolumn{1}{c}{$z_{max}$}&
\multicolumn{1}{c}{R}&\multicolumn{1}{c}{B}&\multicolumn{1}{c}{$\theta$}
\\ \hline
Mrk  421($\zeta=0,1$)   &\  $3.9\times10^2$  &\  $8.0\times10^4$  &\   1.8  &\  $1\times10^{16}$ &\ $6\times10^{16}$ &\ $1\times10^{16}$  &\  0.18 &\ $3^{\circ}$ \\
Mrk  421($\xi=0$)   &\  $4.3\times10^2$  &\  $8.0\times10^4$  &\   1.8  &\  $1\times10^{16}$ &\ $6\times10^{16}$ &\ $1\times10^{16}$  &\  0.24 &\ $3^{\circ}$ \\
Mrk  421($\xi=-1$)   &\  $1.0\times10^2$  &\  $7.0\times10^4$  &\   1.8  &\  $1\times10^{16}$ &\ $6\times10^{16}$ &\ $1\times10^{16}$  &\  0.1 &\ $3^{\circ}$ \\
Mrk  421($\xi=1$)   &\  $1.0\times10^4$  &\  $1.0\times10^5$  &\   1.8  &\  $1\times10^{16}$ &\ $6\times10^{16}$ &\ $1\times10^{16}$  &\  0.9 &\ $3^{\circ}$ \\
0716+714($\zeta=0,1$)  &\  $4.6\times10^4$  &\  $8.0\times10^3$   &\    2.0  &\  $1\times10^{16}$ &\ $6\times10^{16}$ &\ $1\times10^{16}$  &\   1.0 &\ $2^{\circ}$ \\
0716+714($\xi=0$ )  &\  $5.0\times10^4$  &\  $8.0\times10^3$   &\    2.0  &\  $1\times10^{16}$ &\ $6\times10^{16}$ &\ $1\times10^{16}$  &\   1.2 &\ $2^{\circ}$ \\
0716+714($\xi=-1$)  &\  $1.0\times10^4$  &\  $7.0\times10^3$   &\    2.0  &\  $1\times10^{16}$ &\ $6\times10^{16}$ &\ $1\times10^{16}$  &\   0.7 &\ $2^{\circ}$ \\
0716+714($\xi=1$)  &\  $1.0\times10^6$  &\  $1.0\times10^4$   &\    2.0  &\  $1\times10^{16}$ &\ $6\times10^{16}$ &\ $1\times10^{16}$  &\   5.0 &\ $2^{\circ}$ \\
3c279($\xi=0$) &\  $8.0\times10^3$  &\  $1.0\times10^3$   &\    2.0  &\  $1\times10^{16}$ &\ $2\times10^{17}$ &\ $1\times10^{17}$  &\   1.4 &\ $3^{\circ}$ \\
3c279($\xi=-0.5$) &\  $1.0\times10^3$  &\  $9.0\times10^2$   &\    2.0  &\  $1\times10^{16}$ &\ $2\times10^{17}$ &\ $1\times10^{17}$  &\   0.35 &\ $3^{\circ}$ \\
 \hline
\end{tabular}
\end{center}
\end{table}

\clearpage

\end{document}